\documentclass[transmag]{IEEEtran}
\usepackage{latexsym}
\usepackage{graphicx}
\usepackage{amsthm,amsmath, amsfonts, amssymb, amsthm, ragged2e, mathrsfs, balance, xcolor}
\let\bs\boldsymbol
\DeclareMathOperator*{\argmax}{\arg\!\max}
\newtheorem{theorem}{Theorem}

\newtheorem{proposition}{Proposition}

\def\BibTeX{{\rm B\kern-.05em{\sc i\kern-.025em b}\kern-.08em T\kern-.1667em\lower.7ex\hbox{E}\kern-.125emX}}
\newcounter{MYtempeqncnt}
\begin{document}

\title{NOMA-Based Cooperative Relaying with Receive Diversity in Nakagami-$m$ Fading Channels}

\author{Vaibhav~Kumar,~\IEEEmembership{Student~Member,~IEEE,}
        Barry~Cardiff,~\IEEEmembership{Senior~Member,~IEEE,} \\
        and~Mark~F.~Flanagan,~\IEEEmembership{Senior~Member,~IEEE}
\thanks{The authors are with School of Electrical and Electronic Engineering, University College Dublin, Ireland 
(email: vaibhav.kumar@ieee.org, barry.cardiff@ucd.ie, mark.flanagan@ieee.org).}
\thanks{This publication has emanated from research conducted with the financial support of Science Foundation Ireland (SFI) and co-funded under the European Regional Development Fund under Grant Number 13/RC/2077.}
}

\IEEEtitleabstractindextext{\begin{abstract}
Non-orthogonal multiple access (NOMA) is being widely considered as a potential candidate to enhance the spectrum utilization in beyond fifth-generation (B5G) communications. In this paper, we derive closed-form expressions for the ergodic rate and outage probability of a multiple-antenna-assisted NOMA-based cooperative relaying system (CRS-NOMA). We present the performance analysis of the system for two different receive diversity schemes -- selection combining (SC) and maximal-ratio combining (MRC), in Nakagami-$m$ fading. We also evaluate the asymptotic behavior of the CRS-NOMA to determine the slope of the ergodic rate and diversity order. Our results show that in contrast to the existing CRS-NOMA systems, the CRS-NOMA with receive diversity outperforms its orthogonal multiple access (OMA) based counterpart even in the low-SNR regime, by achieving higher ergodic rate. Diversity analysis confirms that the CRS-NOMA achieves full diversity order using both SC and MRC schemes, and this diversity order depends on both the shape parameter $m$ and the number of receive antennas. We also discuss the problem of optimal power allocation for the minimization of the outage probability of the system, and subsequently use this optimal value to obtain the ergodic rate. An excellent match is observed between the numerical and the analytical results, confirming the correctness of the derived analytical expressions. 
\end{abstract}

}

\maketitle

\section{INTRODUCTION}

\IEEEPARstart{W}{ith} the proliferation of wireless communications devices and high-speed data services, the demand for spectrally efficient multiple access technologies are ever-increasing, especially for beyond-fifth-generation (B5G) applications. Non-orthogonal multiple access (NOMA) is by now well-recognized as a potential candidate for efficient spectrum utilization and can also achieve low latency and massive connectivity~\cite{Bhargava},~\cite{Proc_Hanzo}. Although in NOMA, less power is allocated to each individual user as compared to conventional orthogonal multiple access (OMA), the overall spectral efficiency of the system increases because of the more frequent user scheduling and larger bandwidth available for each user. In order to further enhance the performance of NOMA, a user cooperation scheme was proposed in~\cite{Cooperative_NOMA}, where the users with strong channel conditions acted as relays for weak users in orthogonal time slots. 

An intriguing application of NOMA for a decode-and-forward (DF) cooperative relaying system (CRS) was proposed in~\cite{CRS_NOMA}. In contrast to the conventional OMA-based DF relaying where two time slots are needed to deliver a single symbol to the destination, two different symbols were delivered to the destination in two time slots in the system shown in~\cite{CRS_NOMA}. In particular, analytical expressions for the ergodic rate and for near-optimal power allocation over Rayleigh fading were derived in~\cite{CRS_NOMA}. Exact and approximate closed-form expressions for the ergodic rate of CRS-NOMA over Rician fading were presented in~\cite{CRS_NOMA_Rician}. In the CRS-NOMA schemes discussed in~\cite{CRS_NOMA} and~\cite{CRS_NOMA_Rician}, two symbols were independently decoded in two different time slots at the destination. A novel receiver design was proposed for CRS-NOMA in~\cite{Novel_Receiver}, where the destination jointly decoded the two symbols using maximal-ratio combining (MRC), resulting into a (slightly) better ergodic rate as compared to that in~\cite{CRS_NOMA}, at a cost of higher system complexity. In~\cite{CRS_NOMA_AF} the CRS-NOMA with amplify-and-forward (AF) relaying was shown to outperform DF-based CRS-NOMA in terms of the ergodic rate. It is important to note that although the the CRS-NOMA system in~\cite{CRS_NOMA} is shown to outperform CRS-OMA at high transmit SNR ($\rho$), the ergodic rate of CRS-NOMA is worse as compared to that of CRS-OMA in the low-$\rho$ regime. Moreover, the analysis of outage probability of the CRS-NOMA system was not provided in~\cite{CRS_NOMA}.

Multiple-antenna technology is another well-established methodology to increase the spectrum utilization efficiency of a wireless communications system using \emph{spatial diversity}. A multi-antenna assisted relay-based downlink NOMA system was proposed in~\cite{MenGe}, where antenna selection is performed at the transmitter and MRC is employed at the receiver. It was shown in~\cite{MenGe} that the proposed NOMA system enjoys both transmit and receive diversity. A performance analysis of CRS-NOMA with Alamouti space-time block coding (A-STBC) for a $2 \times 1$ multiple-input single-output (MISO) system was presented in~\cite{Kader_STBC}, where the source and the relay were each equipped with two transmit antennas, and the relay and the destination each had a single receive antenna. It can be noted from~\cite{Kader_STBC} that even with an increase in the system complexity (due to the use of A-STBC), the performance of CRS-NOMA was worse in terms of ergodic capacity as compared to that of the CRS-OMA  in the low-$\rho$ regime. 

Motivated by this, a simple but effective multiple-antenna-assisted receiver architecture for CRS-NOMA was proposed in~\cite{SCC}, which enhances the system performance especially in the low-$\rho$ regime. This work considered a CRS-NOMA system where the source and the relay were each equipped with a single transmit antenna, and the relay and the destination were equipped with multiple receive antennas. Selection combining (SC) and MRC were used for signal reception and a performance analysis (in terms of ergodic rate, outage probability and diversity order) in Rayleigh fading was presented in~\cite{SCC}. It was shown that the signal-to-noise ratio (SNR) at which the ergodic rate of CRS-NOMA becomes higher than its OMA-based counterpart decreases with an increase in the number of receive antennas at the destination and the relay.

In this paper, we present a generalization of the results presented in~\cite{SCC} by analyzing the performance of the multiple-antenna-assisted CRS-NOMA in the presence of Nakagami-$m$ fading. The Nakagami-$m$ distribution, which is a mathematically tractable, flexible and generalized fading model, spans the widest range of the amount of fading (AoF) among most of the widely considered multipath fading models. It converges to the Rayleigh, one-sided Gaussian and a nonfading additive white Gaussian noise models as special cases for $m = 1$, $m = 0.5$ and $m \to \infty$, respectively. It can also closely approximate the Hoyt (for $m \leq 1$) and Rice distributions (for $m > 1$). Moreover, the Nakagami-$m$ model offers the best representation of land-mobile propagation, indoor-mobile multipath and ionospheric radio links~\cite{Alouini_Book}. All of these attractive properties justify the motivation for the analysis of multiple-antenna-assisted CRS-NOMA in the presence of Nakagami-$m$ fading. The main contributions of this paper are itemized as follows:
\begin{itemize}
	\item We derive analytical expressions for the ergodic rate of CRS-NOMA over Nakagami-$m$ fading for two different diversity combining schemes -- SC and MRC. This generalizes the results for the ergodic rate derived in~\cite{SCC}. Note that by substituting $m = 1$ into the rate expressions presented in this paper, one can obtain the rate expressions derived in~\cite{SCC}. Furthermore, we also derive an analytical expression for a high-SNR approximation of the ergodic rate for CRS-NOMA, which in turn is used to determine the slope of the ergodic rate in the high-SNR regime\footnote{Note that the analysis of the ergodic rate in the high-SNR regime was not presented in~\cite{SCC}.}.
	\item In~\cite{SCC}, the outage probability for CRS-NOMA was derived by considering the outage of the two different symbols independently. In contrast, in this paper we provide a new definition of the outage probability of CRS-NOMA, where the system is considered to be in outage when \emph{either} or \emph{both} of the symbols are in outage. An analytical expression for the outage probability of CRS-NOMA is derived, and this expression is then used as a system design tool for obtaining the optimal power-allocation coefficient. Note that the analysis in~\cite{SCC} was presented for a fixed value of the power-allocation coefficient.
	\item We also derive the diversity order for the CRS-NOMA and show that the system achieves full diversity order for both diversity combining schemes; this diversity order depends on the shape parameter $m$ as well as the number of receive antennas.
	\item For a given set of system parameters (i.e., number of receive antennas at the destination and the relay, channel statistics, and target data rate of the system), with the help of the analytical expression for the outage probability, we numerically find the optimal power-allocation coefficient that minimizes the outage probability and then use that optimal coefficient in the evaluation of the ergodic rate.
	\item We compare the performance of CRS-NOMA with that of the corresponding OMA system in terms of ergodic rate, and show that the former outperforms the latter, even in the low-SNR regime. We also provide extensive numerical and analytical results to illustrate the effect of the shape parameter $m$ and the number of receive antennas on the performance of CRS-NOMA for both SC and MRC receivers. 
\end{itemize}
The organization of the rest of the paper is as follows: We illustrate the system model in~Section~\ref{Sec-SysMod}. Section~\ref{Sec-SC} deals with the analysis of ergodic rate, outage probability and diversity order of the CRS-NOMA with SC receivers. The performance analysis of CRS-NOMA with MRC receivers is presented in~Section~\ref{Sec-MRC}. In~Section~\ref{Sec-Results} a detailed discussion on the optimal power allocation strategy is presented along with extensive numerical and analytical results. Finally, conclusions are drawn in Section~\ref{Sec-Conclusion}.
\begin{figure}[t]
\centering
\includegraphics[scale = 0.5]{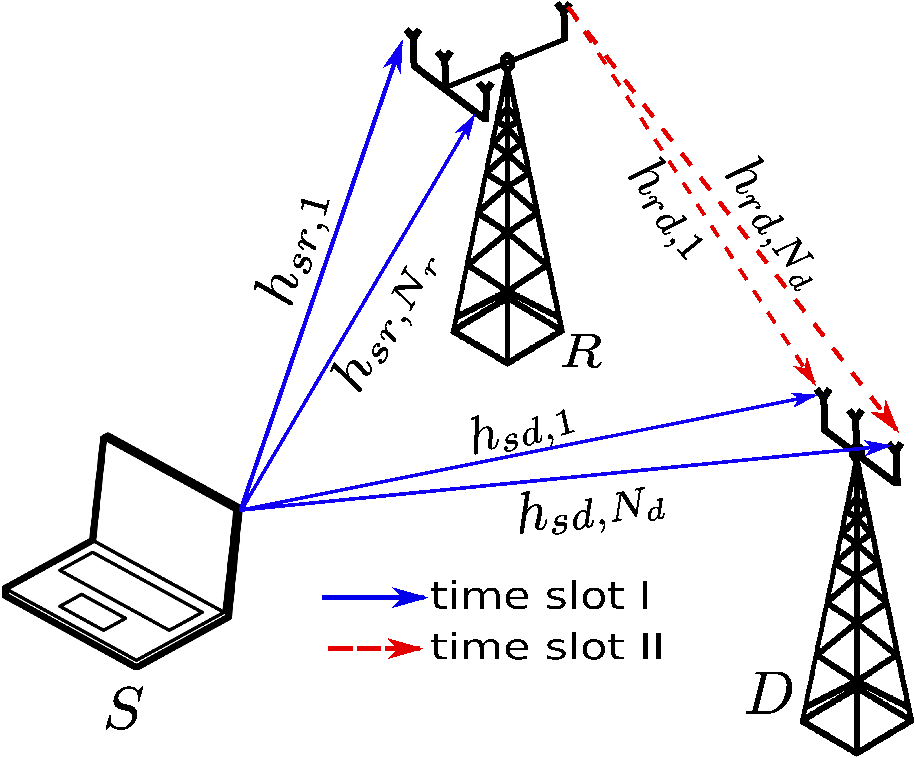}
\caption{System model for CRS-NOMA with multiple receive antennas.}
\label{SysMod}
\end{figure}
\section{System Model} \label{Sec-SysMod}
Consider the three-node system shown in~Fig.~\ref{SysMod} consisting of a source $S$, a relay $R$ and a destination $D$. It is assumed that the system operates in half-duplex mode and that the source and relay each have a single transmit antenna. Moreover, the number of receive antennas at the relay and the destination is denoted by $N_r$ and $N_d$, respectively. We denote the channel coefficient between the transmit antenna of $S$ and the $i^{\mathrm{th}}$ receive antenna of $R$ by $h_{sr, i}$ with $1 \leq i \leq N_r$, and that between the transmit antenna of $S$ and the $j^{\mathrm{th}}$ receive antenna of $D$ by $h_{sd, j}$ with $1 \leq j \leq N_d$. Similarly, the channel coefficient between the transmit antenna of $R$ and the $k^{\mathrm{th}}$ receive antenna of $D$ is denoted by $h_{rd, k}$ with $1 \leq k \leq N_d$. Furthermore, we assume that $h_{sr, i}$, $h_{sd, j}$ and $h_{rd, k}$ are mutually independent and Nakagami-$m$ distributed with shape parameter denoted by $m_{sr} (\forall i)$, $m_{sd} (\forall j)$ and $m_{rd} (\forall k)$, respectively, and mean-square value denoted by $\Omega_{sr}(\forall i)$, $\Omega_{sd} (\forall j)$ and $\Omega_{rd} (\forall k)$, respectively, with $\Omega_{sd} < \Omega_{sr}$. Note that we assume all of the wireless links to follow quasi-static block fading, and we also assume that $m_{sr}, m_{sd}, m_{rd} \in Z_+$, where $\mathbb Z_+$ is the set of positive integers.

For each link, we assume that perfect channel state information (CSI) is available at the receiver side. Also, we assume that only statistical CSI is available at the source $S$ regarding $h_{sr, i}, i \in \{1, 2, \ldots, N_r\}$ and $h_{sd, j}, j \in \{1, 2, \ldots, N_d\}$, and at the relay $R$ regarding $h_{rd, k}, k \in \{1, 2, \ldots, N_d\}$. A similar requirement on CSI knowledge was assumed in~\cite{CRS_NOMA} (except for the addition of multiple antennas).

Denoting by $s_1$ and $s_2$ the unit-energy complex constellation symbols and by $P$ the power budget of the source for each time slot, the symbol transmitted from the source in the first time slot is given by 
\begin{equation}
	\sqrt{a_1 P} s_1 + \sqrt{a_2 P} s_2, \notag 
\end{equation}
where $a_1$ and $a_2$ are the power allocation coefficients with constraints $a_1 > a_2$ and $a_1 + a_2 = 1$. After receiving the signal from $S$, the destination combines the signals and then decodes symbol $s_1$ only by considering the inter-symbol interference from $s_2$ as additional noise, whereas the relay combines the signals and then decodes both symbols (it first decodes $s_1$ and then $s_2$) using successive interference cancellation (SIC)\footnote{Note that in this paper, we assume that no error occurs during SIC. However, in a practical NOMA system, an erroneous decoding of $s_1$ and/or imperfect/outdated CSI will result in imperfect SIC, which will degrade the overall system performance. Therefore, the results presented in this paper for ergodic rate and outage probability  will serve as an upper-bound and a lower-bound, respectively, for those of a practical NOMA system.}. Considering the power budget at the relay to be equal to that of the source, and the SIC at the relay to be perfect, the relay transmits the decoded symbol $s_2$ to the destination using full transmit power $P$ (i.e., $\sqrt{P} s_2$) in the second time slot. In this manner, two different symbols are delivered to the destination in two time slots. In contrast to this, in the conventional OMA-based three-node DF cooperative relaying system, only a single symbol is delivered to the destination in two time slots. 
\sloppy
\section{CRS-NOMA with SC} \label{Sec-SC}
In this section, we provide the performance analysis of CRS-NOMA for the case where selection combining is used for signal reception at both the relay and the destination. The signals received at the relay and the destination in the first time slot are, respectively,
\begin{align*}
	y_{sr, \mathrm{SC}} & = h_{sr, i^*} (\sqrt{a_{1}P} s_{1} + \sqrt{a_{2}P} s_{2}) + n_{sr}, \\	
	y_{sd,\mathrm{SC}} & = h_{sd, j^*} (\sqrt{a_{1}P} s_{1} + \sqrt{a_{2}P} s_{2}) + n_{sd},
\end{align*}
where $i^* = \argmax_{1 \leq i \leq N_{r}}(|h_{sr, i}|)$ and $j^* = \argmax_{1 \leq j \leq N_{d}}(|h_{sd, j}|)$. Moreover, $n_{sr}$ and $n_{sd}$ denote zero-mean complex additive white Gaussian noise (AWGN) with variance $\sigma^{2}$. The relay decodes the symbol $s_1$ with signal-to-interference-plus-noise ratio (SINR) given by $\gamma_{sr, \mathrm{SC}}^{(1)} = g_{sr}a_{1}P/\{g_{sr}a_{2}P + \sigma^{2}\}$, and then applies SIC to decode the symbol $s_2$ with SNR given by $\gamma_{sr, \mathrm{SC}}^{(2)} = g_{sr}a_{2}P/\sigma^{2}$, where $g_{sr} = |h_{sr, i^*}|^{2}$. Similarly, the destination decodes the symbol $s_1$ with SINR given by $\gamma_{sd, \mathrm{SC}} = g_{sd}a_{1}P/\{g_{sd}a_{2}P + \sigma^{2}\}$, where $g_{sd} = |h_{sd, j^*}|^{2}$.

The relay transmits the symbol $s_2$ to the destination in the next time slot. The signal received at the destination is given by
\begin{equation}
	y_{rd, \mathrm{SC}} = h_{rd, k^*} \sqrt{P} s_{2} + n_{rd}, \notag 
\end{equation}
where $k^* = \argmax_{1 \leq k \leq N_{d}} (|h_{rd, k}|)$ and $n_{rd}$ is zero-mean complex AWGN with variance $\sigma^2$. The destination decodes the symbol $s_2$ with SNR given by $\gamma_{rd, \mathrm{SC}} = g_{rd}P/\sigma^{2}$, where $g_{rd} = |h_{rd, k^*}|^{2}$. 

In the following subsection, we present the ergodic rate analysis for CRS-NOMA.
\subsection{Ergodic rate} \label{Sec-SC_Rate}
The ergodic rate for the symbol $s_{1}$ is given by (c.f.~\cite{CRS_NOMA},~\cite{CRS_NOMA_Rician})
\begin{align}
	& \bar{C}_{s_{1}, \mathrm{SC}} \notag \\
	= & \ \mathbb E \left\{ \dfrac{1}{2} \min\left[\log_2(1 + \gamma_{sd, \mathrm{SC}}), \log_2 \left( 1 + \gamma_{sr, \mathrm{SC}}^{(1)} \right) \right] \right\} \notag \\
	= & \ \dfrac{1}{2} \ \mathbb E \left\{ \log_2\left( 1 + \dfrac{a_1 \rho X}{a_2 \rho X + 1}\right) \right\} \notag \\
	= & \ \dfrac{1}{2} \ \mathbb E \left\{ \log_2(1 + \rho X) - \log_2(1 + a_2 \rho X)\right\} \notag \\
	 = & \, \dfrac{1}{2 \ln 2} \left[ \rho \int_{0}^{\infty} \dfrac{1 - F_{X}(x)}{1 + \rho x} dx \right. \notag \\
	 & \hspace{2.5cm}\left. - \rho a_2 \int_{0}^{\infty} \dfrac{1 - F_{X}(x)}{1 + \rho a_2 x} \, dx\right],  \label{C_s1_SC_Integral_Nak}
\end{align}
where $\rho = P/\sigma^{2}$, $X \triangleq \min [ g_{sr}, g_{sd}]$ and $F_{X}(x)$ denotes the cumulative distribution function (CDF) of the random variable~$X$. 
\begin{theorem} \label{Theorem-SC_Cs1}
A closed-form expression for the ergodic rate for symbol $s_1$ in Nakagami-$m$ fading using SC in CRS-NOMA is given by~\eqref{C_s1_SC_Closed_Nak}, shown on the next page, where $\tau = \sum_{\mu = 0}^{m_{sr} - 1} \mu k_{\mu + 1}$, $\omega = \sum_{\nu = 0}^{m_{sd} - 1} \nu l_{\nu + 1}$, $\Psi_{k_0, l_0}~=~\frac{(N_r - k_0) m_{sr}}{\Omega_{sr}} + \frac{(N_d - l_0) m_{sd}}{\Omega_{sd}}$, $\Gamma(\cdot)$ denotes the Gamma function and $\Gamma[\cdot, \cdot]$ denotes the upper-incomplete Gamma function.
\end{theorem}
\begin{figure*}[!t]
\normalsize
\setcounter{MYtempeqncnt}{\value{equation}}
\begin{align}
		\bar C_{s_1, \mathrm{SC}} = \dfrac{1}{2 \ln 2} & \left[ \sum_{\substack{ k_0 + k_1 + \cdots + k_{m_{sr}} = N_r \\ k_0 \neq N_r}} \sum_{\substack{ l_0 + l_1 + \cdots + l_{m_{sd}} = N_d \\ l_0 \neq N_d}} \binom{N_r}{k_0, k_1, \ldots, k_{m_{sr}}} \binom{N_d}{l_0, l_1, \ldots, l_{m_{sd}}} (-1)^{N_r + N_d - k_0 - l_0} \right. \notag \\
		& \times \left\{ \prod_{\mu = 0}^{m_{sr} - 1} \left( \dfrac{1}{\mu !}\right)^{k_{\mu + 1}}\right\} \left( \dfrac{m_{sr}}{\Omega_{sr}}\right)^{\tau} \left\{ \prod_{\nu = 0}^{m_{sd} - 1} \left( \dfrac{1}{\nu !}\right)^{l_{\nu + 1}}\right\} \left( \dfrac{m_{sd}}{\Omega_{sd}}\right)^{\omega} \dfrac{\Gamma(\tau + \omega + 1)}{\rho^{\tau + \omega}} \left\{ \exp \left( \dfrac{\Psi_{k_0, l_0}}{\rho}\right) \right. \notag \\
		& \hspace{4.5cm}\left. \times \Gamma\left[ -\tau-\omega, \dfrac{\Psi_{k_0, l_0}}{\rho}\right] - \dfrac{1}{a_2^{\tau + \omega}}\exp \left( \dfrac{\Psi_{k_0, l_0}}{\rho a_2} \right) \Gamma \left[ -\tau -\omega, \dfrac{\Psi_{k_0, l_0}}{\rho a_2}\right] \right\} \Bigg].\label{C_s1_SC_Closed_Nak}
\end{align}
\setcounter{equation}{\value{MYtempeqncnt}}
\hrulefill
\end{figure*}
\addtocounter{equation}{1}
\begin{proof}
	See Appendix~\ref{Proof-SC_Cs1}.
\end{proof}

Similarly, the ergodic rate for the symbol $s_2$ is given by (c.f.~\cite{CRS_NOMA},~\cite{CRS_NOMA_Rician})
\begin{align}
	& \bar C_{s_2, \mathrm{SC}} \notag \\
	= & \ \mathbb E \left\{ \dfrac{1}{2} \min \left[ \log_2 \left( 1 + \gamma_{sr, \mathrm{SC}}^{(2)} \right), \log_2 \left( 1 + \gamma_{rd, \mathrm{SC}} \right)\right] \right\} \notag \\
	= & \ \dfrac{1}{2} \mathbb E \left\{ \log_2 \left( 1 + \rho \min[g_{sr} a_2, g_{rd}] \right)\right\} \notag \\
	= & \ \dfrac{\rho}{2 \ln 2} \int_0^\infty \dfrac{1 - F_Y(x)}{1 + \rho x} \ \mathrm dx, \label{C_s2_SC_Integral_Nak}
\end{align}
where $Y \triangleq \min [ g_{sr} a_2, g_{rd} ]$.
\begin{proposition}
A closed-form expression for the ergodic rate for symbol $s_2$ for the CRS-NOMA using SC in Nakagami-$m$ fading is given by~\eqref{C_s2_SC_Closed_Nak}, shown on the next page, where $\alpha_{k_0, q_0} = \tfrac{(N_r - k_0) m_{sr}}{a_2 \Omega_{sr}}+\tfrac{(N_d - q_0) m_{rd}}{\Omega_{rd}} $ and $\upsilon~=~\sum_{\varepsilon = 0}^{m_{rd} - 1}\varepsilon q_{\varepsilon + 1}$.
\end{proposition}
\begin{figure*}[!t]
\normalsize
\setcounter{MYtempeqncnt}{\value{equation}}
\begin{align}
		\bar C_{s_2, \mathrm{SC}} = & \dfrac{1}{2\ln 2}\sum_{\substack{{k_0 + k_1 + \cdots + k_{m_{sr}} = N_r} \\ k_0 \neq N_r}}  \sum_{\substack{{q_0 + q_1 + \cdots + q_{m_{rd}} = N_d} \\ q_0 \neq N_d}} \binom{N_r}{k_0, k_1, \ldots, k_{m_{sr}}}  \binom{N_d}{q_0, q_1, \ldots, q_{m_{rd}}} (-1)^{N_r + N_d - k_0 - q_0} \notag \\
		& \times \left[ \prod_{\mu = 0}^{m_{sr} - 1} \left( \dfrac{1}{\mu!}\right)^{k_{\mu + 1}}\right] \left( \dfrac{m_{sr}}{a_2 \Omega_{sr}}\right)^{\tau}\left[ \prod_{\varepsilon = 0}^{m_{rd} - 1} \left( \dfrac{1}{\varepsilon!}\right)^{q_{\varepsilon + 1}}\right]  \left( \dfrac{m_{rd}}{\Omega_{rd}} \right)^{\upsilon} \dfrac{\Gamma(\tau + \upsilon + 1)}{\rho^{\tau + \upsilon}}\! \exp\! \left(\!\! \dfrac{\alpha_{k_0, q_0}}{\rho}\!\!\right)  \!\Gamma \!\left[\! -\tau\!-\!\upsilon, \!\dfrac{\alpha_{k_0, q_0}}{\rho}\right]. \label{C_s2_SC_Closed_Nak}
\end{align}
\setcounter{equation}{\value{MYtempeqncnt}}
\hrulefill
\end{figure*}
\addtocounter{equation}{1}
\begin{proof}
Similar to the arguments in Appendix~\ref{Proof-SC_Cs1} and using a transformation of random variables, for $m_{rd} \in \mathbb Z_+$ we have 
\begin{align}
	& 1 - F_Y\!(x) =  \sum_{\substack{{k_0 + \cdots + k_{m_{sr}} = N_r} \\ k_0 \neq N_r}}  \sum_{\substack{{q_0 + \cdots + q_{m_{rd}} = N_d} \\ q_0 \neq N_d}} \notag \\
	& \times \binom{N_r}{\!k_0, k_1, \ldots, k_{m_{sr}\!}} \binom{N_d}{\!q_0, q_1, \ldots, q_{m_{rd}\!}} (-1)^{N_r + N_d - k_0 - q_0}  \notag \\
	& \times \left[ \prod_{\mu = 0}^{m_{sr} - 1} \left( \dfrac{1}{\mu!}\right)^{k_{\mu + 1}}\right] \left( \dfrac{m_{sr}}{a_2 \Omega_{sr}} \right)^{\tau} \left[\prod_{\varepsilon = 0}^{m_{rd} - 1} \left( \dfrac{1}{\varepsilon!}\right)^{q_{\varepsilon + 1}}\right]\notag \\
	& \hspace{2.3cm}\times  \left( \dfrac{m_{rd}}{\Omega_{rd}} \right)^{\upsilon}x^{\tau + \upsilon} \exp \left(-\alpha_{k_0, q_0}x \right). \label{1-FmathscrY}
\end{align}
Using \eqref{C_s2_SC_Integral_Nak}, \eqref{1-FmathscrY} and~\cite[eqn.~(3.383-10),~p.~348]{Grad}, the closed-form expression for~$\bar C_{s_2, \mathrm{SC}}$ becomes equal to~\eqref{C_s2_SC_Closed_Nak}.
\end{proof}

Therefore, using \eqref{C_s1_SC_Closed_Nak} and \eqref{C_s2_SC_Closed_Nak}, the ergodic rate for the CRS-NOMA using SC in Nakagami-$m$ fading is given by
\begin{align}
	\bar C_{\mathrm{SC}} = \bar C_{s_1, \mathrm{SC}} + \bar C_{s_2, \mathrm{SC}}. \label{C_sum_SC}
\end{align}
It can be shown with the help of some algebraic manipulations that for $m_{sr} = m_{sd} = m_{rd} = 1$, \eqref{C_sum_SC} reduces to~\cite[eqn.~(6)]{SCC}. Furthermore, for $m_{sr} = m_{sd} = m_{rd} = 1$ and $N_r = N_d = 1$, \eqref{C_sum_SC} reduces to~\cite[eqn.~(14)]{CRS_NOMA}.

Contrary to the NOMA-based system, in CRS-OMA the source $S$ transmits the symbol $s_1$ to both relay and destination in the first time slot. The signals received at the corresponding nodes are given by
\begin{align*}
	y_{sr, \mathrm{SC-OMA}} = & \ h_{sr, i^*} \sqrt{P} s_1 + n_{sr}, \\
	y_{sd, \mathrm{SC-OMA}} = & \ h_{sd, j^*} \sqrt{P} s_1 + n_{sd}.
\end{align*}
After receiving the signal from the source, the relay decodes the symbol $s_1$ and re-encodes and forwards this symbol to the destination in the next time slot. The signal received at the destination is given by 
\begin{align*}
	y_{rd, \mathrm{SC-OMA}} = h_{rd,k^*} \sqrt{P} s_1 + n_{rd}. 
\end{align*}
Therefore, assuming that the destination applies MRC on the two copies of $s_1$ (received in two orthogonal time slots), the ergodic rate of CRS-OMA with SC is given by (c.f.~\cite{Laneman})
\begin{align*}
	\bar C_{\mathrm{SC-OMA}} = 0.5 \mathbb E_W \left[ \log_2(1 + W \rho)\right],
\end{align*}
where $W \triangleq \min\{g_{sr}, g_{sd} + g_{rd}\}$. Since the focus of this paper is on the NOMA-based systems, we do not present a closed-form analysis for CRS-OMA.
%
%
\subsection{High-SNR approximation of the ergodic rate} \label{Sec-SC_HighSNR}
Although the analytical expressions in~\eqref{C_s1_SC_Closed_Nak} and~\eqref{C_s2_SC_Closed_Nak} represent the exact ergodic rate for $s_1$ and $s_2$, respectively, it is difficult to draw insights from these expressions. Therefore, in this section, we derive a high-SNR approximation of the ergodic rate of $s_1$ and $s_2$. 

From~\eqref{C_s1_SC_Integral_Nak}, it is straightforward to show that for $\rho~\gg~1$, 
\begin{align}
	\bar C_{s_1, \mathrm{SC}} = & \ \dfrac{1}{2} \mathbb E \left\{ \log_2 \left( 1 + \dfrac{a_1 \rho X}{a_2 \rho X + 1}\right) \right\} \notag \\
	 \approx & \ \dfrac{1}{2} \log_2 \left( 1 + \dfrac{a_1}{a_2}\right). \label{C_s1_SC_HighSNR}
\end{align}
It can be noted from~\eqref{C_s1_SC_HighSNR} that at high SNR, the ergodic rate of $s_1$ becomes (almost) constant (w.r.t. $\rho$) and depends only on the ratio of the power allocation coefficients $a_1$ and $a_2$. On the other hand, for the case of $s_2$ at $\rho \gg 1$, it can be deduced from~\eqref{C_s2_SC_Integral_Nak} that
\begin{align}
	& \bar C_{s_2 \mathrm{SC}} \approx \dfrac{1}{2} \mathbb E \left\{ \log_2 (\rho Y)\right\} = \dfrac{1}{2} \log_2 \rho + \dfrac{1}{2} \mathbb E \left\{ \log_2 Y \right\} \notag \\
	= & \ \dfrac{1}{2} \log_2 \rho + \dfrac{1}{2} \int_0^\infty \log_2(x) f_Y(x) \mathrm dx, \label{C_s2_SC_HighSNR_integral}
\end{align}
where $f_Y(x)$ denotes the PDF of the random variable $Y$. Using~\eqref{1-FmathscrY}, the expression for $f_Y(x)$ can be given by~\eqref{fY}, shown on the next page.

\begin{figure*}[!t]
\normalsize
\setcounter{MYtempeqncnt}{\value{equation}} 
\begin{align}
	& f_Y(x) = \dfrac{\mathrm dF_Y(x)}{\mathrm dx} = \sum_{\substack{{k_0 + \cdots + k_{m_{sr}} = N_r} \\ k_0 \neq N_r}}  \sum_{\substack{{q_0 + \cdots + q_{m_{rd}} = N_d} \\ q_0 \neq N_d}} \binom{N_r}{k_0, k_1, \ldots, k_{m_{sr}}} \binom{N_d}{q_0, q_1, \ldots, q_{m_{rd}}} (-1)^{N_r + N_d - k_0 - q_0}  \notag \\
	& \times \left[ \prod_{\mu = 0}^{m_{sr} - 1} \left( \dfrac{1}{\mu!}\right)^{k_{\mu + 1}}\right] \left( \dfrac{m_{sr}}{a_2 \Omega_{sr}} \right)^{\tau} \left[ \prod_{\varepsilon = 0}^{m_{rd} - 1} \left( \dfrac{1}{\varepsilon!} \right)^{ q_{\varepsilon + 1}}\right] \left( \dfrac{m_{rd}}{\Omega_{rd}} \right)^{\upsilon} \exp(-\alpha_{k_0, q_0} x) \left[ \alpha_{k_0, q_0} x^{\tau + \upsilon} -(\tau + \upsilon) x^{\tau + \upsilon - 1} \right]. \label{fY}
\end{align}
\setcounter{equation}{\value{MYtempeqncnt}}
\hrulefill
\end{figure*}
\addtocounter{equation}{1}
Substituting the expression for $f_Y(x)$ from~\eqref{fY} into~\eqref{C_s2_SC_HighSNR_integral} and solving the integral using~\cite[eqn.~(4.352-1),~p.~573]{Grad}, a high-SNR approximation of $\bar C_{s_2, \mathrm{SC}}$ is given by~\eqref{C_s2_SC_HighSNR}, shown on the next page, where $\psi(\cdot)$ is the Euler psi function or digamma function (defined as the logarithmic derivative of the Gamma function). Therefore, a high-SNR approximation of $\bar C_{\mathrm{SC}}$ can be obtained by adding~\eqref{C_s1_SC_HighSNR} and~\eqref{C_s2_SC_HighSNR}.

\begin{figure*}[!t]
\normalsize
\setcounter{MYtempeqncnt}{\value{equation}}
\begin{align}
	\bar C_{s_2, \mathrm{SC}} \approx & \ \dfrac{1}{2} \log_2 \rho + \dfrac{1}{2 \ln 2} \sum_{\substack{{k_0 + \cdots + k_{m_{sr}} = N_r} \\ k_0 \neq N_r}}  \sum_{\substack{{q_0 + \cdots + q_{m_{rd}} = N_d} \\ q_0 \neq N_d}} \binom{N_r}{k_0, k_1, \ldots, k_{m_{sr}}} \binom{N_d}{q_0, q_1, \ldots, q_{m_{rd}}}  \notag \\
	& \times  (-1)^{N_r + N_d - k_0 - q_0} \left[ \prod_{\mu = 0}^{m_{sr} - 1} \left( \dfrac{1}{\mu!}\right)^{k_{\mu + 1}}\right] \left( \dfrac{m_{sr}}{a_2 \Omega_{sr}} \right)^{\tau} \left[ \prod_{\varepsilon = 0}^{m_{rd} - 1} \left( \dfrac{1}{\varepsilon!} \right)^{ q_{\varepsilon + 1}}\right] \left( \dfrac{m_{rd}}{\Omega_{rd}} \right)^{\upsilon} \dfrac{1}{\alpha_{k_0, q_0}^{\tau + \upsilon}} \notag \\
	& \times [ \Gamma(\tau + \upsilon + 1) \{ \psi(\tau + \upsilon + 1) - \ln \alpha_{k_0, q_0} \} - (\tau + \upsilon) \Gamma(\tau + \upsilon) \left\{ \psi(\tau + \upsilon) - \ln \alpha_{k_0, q_0} \right\} ]. \label{C_s2_SC_HighSNR}
\end{align}
\setcounter{equation}{\value{MYtempeqncnt}}
\hrulefill
\end{figure*}
\addtocounter{equation}{1}

%
%
It has already been noted from~\eqref{C_s1_SC_HighSNR} that the ergodic rate of $s_1$ becomes approximately independent of $\rho$ at high SNR. On the other hand, the first term $0.5 \log_2 \rho$ on the right-hand side of~\eqref{C_s2_SC_HighSNR} exhibits a positive constant slope when plotted against $\rho$ in dB, while the terms thereafter are independent of $\rho$. Therefore, it is straightforward to conclude that at high SNR, $\bar C_{\mathrm{SC}}$ grows as $0.5 \log_2 \rho + \mathscr O(1)$, where $\mathscr O(\cdot)$ denotes the Landau symbol.

It is worth noticing that this result is in line with~\cite{MIMO_Goldsmith}, where it has been established that for an $M_T \times M_R$ (here $M_T$ and $M_R$ denote the number of transmit and receive antennas, respectively) multiple-input multiple-output (MIMO) system, the capacity grows as $\min(M_T, M_R) \log_2 \rho + \mathscr O(1)$, as $\rho \to \infty$. Note that in our system, the number of transmit antennas is equal to one in both time slots (i.e., $\min (M_T, M_R)$ for our system is equal to 1), and the factor of $0.5$ (in the term $0.5 \log_2 \rho$) appears due to the relaying protocol.
%
%
\subsection{Outage probability} \label{Sec-SC_Outage}
Assuming the target data rate for both symbols to be the same, denoted by $R$, and defining $\theta \triangleq 2^{2R} - 1$, the non-outage event for the system is defined as the event when the symbol $s_1$ is successfully decoded in the first time slot at the destination, the symbols $s_1$ and $s_2$ are successfully decoded in the first time slot at the relay, and the symbol $s_2$ is successfully decoded in the second time slot at the destination. Using this definition, the outage probability for the CRS-NOMA with SC is defined as
\begin{align}
	 & P_{\mathrm{SC}} \triangleq \! 1 - \Pr (\gamma_{sd, \mathrm{SC}} \geq \theta) \Pr \left( \gamma_{sr, \mathrm{SC}}^{(1)} \geq \theta, \gamma_{sr, \mathrm{SC}}^{(2)} \geq \theta \right) \notag \\
	 \times & \Pr \left( \gamma_{rd, \mathrm{SC}} \geq \theta \right) =  1 - \Pr \left( \dfrac{a_1 \rho g_{sd}}{a_2 \rho g_{sd} + 1} \geq \theta \right) \notag \\
	\times & \Pr \left( \dfrac{a_1 \rho g_{sr}}{a_2 \rho g_{sr} + 1} \geq \theta, a_2 \rho g_{sr} \geq \theta \right) \Pr \left( g_{rd} \rho \geq \theta \right) \notag \\
	= & 1 - \Pr(g_{sd} \geq \Delta_1) \ \Pr(g_{sr} \geq \Delta_2) \ \Pr(g_{rd} \geq \Delta_3) \notag \\
	= & 1 - \left\{1 - F_{g_{sd}}(\Delta_1) \right\} \left\{1 - F_{g_{sr}}(\Delta_2) \right\} \left\{1 - F_{g_{rd}}(\Delta_3) \right\}, \label{Outage_SC_Def}
\end{align} 
where $\Delta_1 \triangleq \theta/\{\rho (a_1 - a_2 \theta)\}$, $\Delta_2 \triangleq \max\{\theta/\{\rho (a_1 - a_2 \theta)\}, \theta/(a_2 \rho)\}$ and $\Delta_3 \triangleq \theta/\rho$. It is important to note that~\eqref{Outage_SC_Def} holds good only for the case when $a_1 > a_2 \theta$, or equivalently, $a_2 < 1/(1+\theta)$; otherwise, both the relay and destination will fail to decode $s_1$ in the first time slot and the outage probability of the system will always be 1. A similar phenomenon was reported in~\cite{RelaySelection}. Using~\eqref{F_gSR_SC} and~\eqref{Outage_SC_Def}, an analytical expression for the outage probability of CRS-NOMA with SC is given by~\eqref{Outage_SC_Closed}, shown on the next page.
\begin{figure*}[!t]
\normalsize
\setcounter{MYtempeqncnt}{\value{equation}}
\begin{align}
		P_{\mathrm{SC}} =  1 - & \left\{ 1 - \left[ \dfrac{\gamma \left( m_{sd}, \frac{m_{sd}}{\Omega_{sd}}\Delta_1\right)}{\Gamma(m_{sd}) } \right]^{N_d}\right\} \left\{ 1 - \left[ \dfrac{ \gamma \left( m_{sr}, \frac{m_{sr}}{\Omega_{sr}}\Delta_2 \right) }{\Gamma(m_{sr})}\right]^{N_r}\right\} \left\{ 1 - \left[ \dfrac{ \gamma \left( m_{rd}, \frac{m_{rd}}{\Omega_{rd}}\Delta_3\right) }{\Gamma(m_{rd}) }  \right]^{N_d}\right\}. \label{Outage_SC_Closed}
\end{align}
\setcounter{equation}{\value{MYtempeqncnt}}
\hrulefill
\end{figure*}
\addtocounter{equation}{1}
%
%
\subsection{Diversity analysis} \label{Sec-SC_Diversity}
In order to obtain insights on the asymptotic performance of the system, we present the diversity analysis of CRS-NOMA with SC in this subsection. With some algebraic manipulation,~\eqref{Outage_SC_Def} can be written as 
\begin{align}
	& P_{\mathrm{SC}} =  F_{g_{sd}}(\Delta_1) + F_{g_{sr}}(\Delta_2) + F_{g_{rd}}(\Delta_3) \notag \\
	& - F_{g_{sd}}(\Delta_1) F_{g_{sr}}(\Delta_2) - F_{g_{sr}}(\Delta_2) F_{g_{rd}}(\Delta_3) \notag \\
	& - F_{g_{rd}}(\Delta_3) F_{g_{sd}}(\Delta_1) + F_{g_{sd}}(\Delta_1) F_{g_{sr}}(\Delta_2) F_{g_{rd}}(\Delta_3). \label{Outage_SC_Def_Expand}
\end{align}
Define the function
\begin{align}
	\mathcal T(m, \Omega, x) \triangleq \dfrac{1}{\Gamma(m)} \ \gamma \left( m, \dfrac{m x}{\Omega}\right). \label{T_function}
\end{align}
Using the series expansion of the exponential and the lower-incomplete Gamma functions~\cite[eqn.~(8.11.4), p.~180]{NIST}, we have 
\begin{align}
	\mathcal T(m, \Omega, x) = & \ \sum_{\beta = 0}^{\infty} \sum_{\delta = 0}^{\infty} \dfrac{(-1)^{\beta}}{\beta ! \ \Gamma(m + \delta + 1)} \left( \dfrac{m x}{\Omega} \right)^{m + \beta + \delta} \notag \\
	= & \ \dfrac{1}{\Gamma(m)}\left( \dfrac{m}{\Omega}\right)^m x^m + \mathscr O \left( x^{m + 1}\right) \notag \\
	\triangleq & \ \mathfrak C(m, \Omega) \ x^m + \mathscr O \left( x^{m + 1}\right). \label{T_function_expansion}
\end{align}
Using the multinomial theorem and~\eqref{F_gSR_SC}, the expression for $F_{g_{sr}}(\Theta_1)$ can be represented as 
\begin{align}
	& F_{g_{sd}}(\Delta_1) = \left[ \mathcal T(m_{sd}, \Omega_{sd}, \Delta_1) \right]^{N_d} \notag \\
	= & \ \left[ \mathfrak C(m_{sd}, \Omega_{sd}) \right]^{N_d} \ \Delta_1^{m_{sd} N_d} + \mathscr O \left( \Delta_1^{m_{sd} N_d + 1}\right) \notag \\
	= & \ \left[ \mathfrak C(m_{sd}, \Omega_{sd}) \left( \dfrac{ \theta }{a_1 - a_2 \theta } \right)^{m_{sd}} \right]^{N_d}  \rho^{-m_{sd} N_d} \notag \\
	& \hspace{3.5cm} + \mathscr O  \left( \rho^{-(m_{sd} N_d + 1)}\right). \label{F_gSD_SC_Order}
\end{align}
It is clear from~\eqref{F_gSD_SC_Order} that $F_{g_{sd}}(\Delta_1)$ decays as $\rho^{-m_{sd}N_d}$ for $\rho \to \infty$. Similarly, it can be shown that $F_{g_{sr}}(\Delta_2)$ and $F_{g_{rd}}(\Delta_3)$ decay as $\rho^{-m_{sr}N_r}$ and $\rho^{-m_{rd}N_d}$, respectively, for $\rho \to \infty$. Therefore, it is straightforward to conclude using~\eqref{Outage_SC_Def_Expand} that the diversity order of CRS-NOMA with SC is $\min\{m_{sd}N_d, m_{sr}N_r, m_{rd}N_d\}$. 

In the next section, we analyze the performance of CRS-NOMA with MRC. 
%
%
\section{CRS-NOMA with MRC} \label{Sec-MRC}
The signals received (after applying MRC) in the first time slot at the relay and the destination are respectively given by
\begin{align}
	& y_{sr, \mathrm{MRC}} = \bs{h}_{sr}^{H} \, \left( \bs{h}_{sr} \left(\sqrt{a_{1}P} s_{1} + \sqrt{a_{2}P}s_{2}\right) + \bs{n}_{sr}\right), \notag \\ 
	& y_{sd, \mathrm{MRC}} = \bs{h}_{sd}^{H} \, \left( \bs{h}_{sd} \left(\sqrt{a_{1}P} s_{1} + \sqrt{a_{2}P}s_{2}\right) + \bs{n}_{sd} \right), \notag 
\end{align}
where $\bs{h}_{sr} = [h_{sr, 1}\, h_{sr, 2}\, \cdots \, h_{sr, N_{r}}]^{T} \in \mathbb{C}^{N_{r} \times 1}$, $\bs{h}_{sd} = [h_{sd, 1}\, h_{sd, 2}\, \cdots \, h_{sd, N_{d}}]^{T} \in \mathbb{C}^{N_{d} \times 1}$, $\bs{n}_{sr} = [n_{sr, 1}\, n_{sr, 2}$ $\, \cdots \, n_{sr, N_{r}}]^{T} \in~\mathbb{C}^{N_{r} \times 1}$, $\bs{n}_{sd} = [n_{sd, 1}\, n_{sd, 2}\, \cdots \, n_{sd, N_{d}}]^{T} \in \mathbb{C}^{N_{d} \times 1}$, $(\cdot)^{H}$ is the Hermitian operator and $(\cdot)^T$ is the transpose operator. The elements in vectors $\bs{h}_{sr}$ and $\bs{h}_{sd}$ are independent and distributed according to the Nakagami-$m$ distribution with shape parameter $m_{sr}$ and $m_{sd}$, respectively, and mean-square value $\Omega_{sr}$ and $\Omega_{sd}$, respectively. The elements in $\bs{n}_{sr}$ and $\bs{n}_{sd}$ are independent and distributed according to $\mathcal{CN}(0, \sigma^2)$.

The relay decodes the symbol $s_1$ with SINR $\gamma_{sr, \mathrm{MRC}}^{(1)} = \mathfrak g_{sr}a_{1}P/\{\mathfrak g_{sr}a_{2}P + \sigma^{2}\}$ and then applies SIC to decode the symbol $s_2$ with SNR $\gamma_{sr, \mathrm{MRC}}^{(2)} = \mathfrak g_{sr}a_{2}P/\sigma^{2}$, where $\mathfrak g_{sr} = \sum_{i = 1}^{N_r}|h_{sr, i}|^{2}$. Similarly, the destination decodes the symbol $s_1$ with SINR $\gamma_{sd, \mathrm{MRC}} = \mathfrak g_{sd} a_{1}P/\{\mathfrak g_{sd}a_{2}P + \sigma^{2}\}$, where $\mathfrak g_{sd} = \sum_{j = 1}^{N_d}|h_{sd, j}|^{2}$.
\sloppy
The relay transmits the symbol $s_2$ to the destination. The signal received at the destination is given by
\begin{align}
	y_{rd, \mathrm{MRC}} = \bs{h}_{rd}^H \left(  \bs{h}_{rd} \sqrt{P}s_2 + \bs{n}_{rd} \right), \notag 
\end{align}
where $\bs{h}_{rd} = \left[h_{rd, 1} \, h_{rd, 2}\, \cdots\, h_{rd, N_d}\right]^T \in \mathbb C^{N_d \times 1}$ with independent elements each of which is Nakagami-$m$ distributed with shape parameter $m_{rd}$ and mean-square value $\Omega_{rd}$, and $\bs{n}_{rd} = [n_{rd, 1}\, n_{rd, 2}$ $\, \cdots \, n_{rd, N_{d}}]^{T} \in~\mathbb{C}^{N_{d} \times 1}$ contains independent elements each distributed according to $\mathcal{CN}(0, \sigma^2)$. The destination decodes the symbol $s_2$ with SNR $\gamma_{rd, \mathrm{MRC}} = \mathfrak g_{rd}P/\sigma^{2}$, where $\mathfrak g_{rd} = \sum_{k = 1}^{N_d}|h_{rd, k}|^2$.

In the following subsection, we present the ergodic rate analysis for CRS-NOMA with MRC. 
%
%
\subsection{Ergodic rate} \label{Sec-MRC_Rate}
The ergodic rate for symbol $s_1$ is given by
\begin{align}
	& \bar C_{s_1, \mathrm{MRC}} = \dfrac{1}{2} \mathbb E \left[ \log_2 (1 + \rho \mathcal X) - \log_2(1 + a_2 \rho \mathcal X)\right] \notag \\
	= & \ \dfrac{1}{2 \ln 2} \left[ \rho \int_{0}^{\infty} \dfrac{1 - F_{\mathcal X}(x)}{1 + \rho x} dx \right. \notag \\
	& \hspace{2.5cm}\left. - \rho a_2 \int_{0}^{\infty} \dfrac{1 - F_{\mathcal X}(x)}{1 + \rho a_2 x} \, dx\right],  \label{C_s1_MRC_Def}
\end{align}
where $\mathcal X \triangleq \min\{\mathfrak g_{sr}, \mathfrak g_{sd}\}$. 
\begin{theorem} \label{Theorem-MRC_Cs1}
In the case of CRS-NOMA with MRC in Nakagami-$m$ fading, an analytical expression for the ergodic rate for the symbol $s_1$ is given by~\eqref{C_s1_MRC_Closed}, shown on the next page, where $\Phi = (m_{sr}/\Omega_{sr}) + (m_{sd}/\Omega_{sd})$.
\end{theorem}
\begin{figure*}[!t]
\normalsize
\setcounter{MYtempeqncnt}{\value{equation}}
\begin{align}
		& \bar C_{s_1, \mathrm{MRC}} = \dfrac{1}{2\ln 2} \sum_{\mu =0}^{ m_{sr}N_r - 1}\sum_{\nu = 0}^{m_{sd}N_d - 1} \dfrac{m_{sr}^{\mu} m_{sd}^{\nu} \Gamma(\mu + \nu + 1)}{\Omega_{sr}^{\mu} \Omega_{sd}^{\nu} \mu! \nu! \rho^{\mu + \nu}} \left\{\! \exp \left(\!\dfrac{\Phi}{\rho} \!\right)\! \Gamma \left[\! -\mu\! -\! \nu, \dfrac{\Phi}{\rho} \right]\! -\! \dfrac{1}{a_2^{\mu + \nu}} \exp \left(\!\dfrac{\Phi}{\rho a_2} \!\right) \Gamma \left[\!-\mu \!-\! \nu, \dfrac{\Phi}{\rho a_2} \right] \right\}. \label{C_s1_MRC_Closed}
\end{align}
\setcounter{equation}{\value{MYtempeqncnt}}
\hrulefill
\end{figure*}
\addtocounter{equation}{1}

\begin{proof}
	See~Appendix~\ref{Proof-MRC_Cs1}.
\end{proof}
Similarly, for the symbol $s_2$, the ergodic rate is given by
\begin{align}
	\bar C_{s_2, \mathrm{MRC}} = & \ \dfrac{1}{2} \mathbb E \left[ \log_2 (1 + \rho \ \min\{\mathfrak g_{sr} a_2, \mathfrak g_{rd}\}) \right] \notag \\
	= & \ \dfrac{\rho}{2 \ln 2} \int_0^\infty \dfrac{1 - F_{\mathcal Y}(x)}{1 + \rho x} \ \mathrm dx, \label{C_s2_MRC_Def}
\end{align}
where $\mathcal Y \triangleq \min\{\mathfrak g_{sr}a_2, \mathfrak g_{rd}\}$.
\begin{proposition}
In the case of CRS-NOMA with MRC in Nakagami-$m$ fading, an analytical expression for the ergodic rate for the symbol $s_2$ is given by~\eqref{C_s2_MRC_Closed}, shown on the next page, where $\Xi = \tfrac{m_{sr}}{\Omega_{sr}a_2} + \tfrac{m_{rd}}{\Omega_{rd}}$.
\end{proposition}
\begin{figure*}[!t]
\normalsize
\setcounter{MYtempeqncnt}{\value{equation}}
\begin{align}
		& \bar C_{s_2, \mathrm{MRC}} = \dfrac{1}{2 \ln 2} \sum_{\mu = 0}^{m_{sr}N_r - 1}\sum_{\nu = 0}^{m_{rd}N_d - 1} \dfrac{m_{sr}^{\mu} m_{rd}^{\nu}}{\Omega_{sr}^\mu \Omega_{rd}^\nu a_2^{\mu} \mu! \nu! }  \dfrac{\Gamma(\mu + \nu + 1)}{\rho^{\mu + \nu}}\exp \left( \dfrac{\Xi}{\rho}\right) \Gamma \left[ -\mu - \nu, \dfrac{\Xi}{\rho}\right]. \label{C_s2_MRC_Closed}
\end{align}
\setcounter{equation}{\value{MYtempeqncnt}}
\hrulefill
\end{figure*}
\addtocounter{equation}{1}

\begin{proof}
Following the arguments in Appendix~\ref{Proof-MRC_Cs1} and with the help of a transformation of random variables, we have
\begin{align}
	& 1 - F_{\mathcal Y}(x) \notag \\
	= & \ \exp( -\Xi x) \sum_{\mu = 0}^{m_{sr} N_r - 1} \sum_{\nu = 0}^{m_{rd}N_d - 1} \dfrac{m_{sr}^\mu m_{rd}^\nu x^{\mu + \nu}}{a_2^{\mu}\Omega_{sr}^\mu \Omega_{rd}^\nu \mu! \nu!}, \label{1-FmathfrakY}
\end{align}
Using \eqref{C_s2_MRC_Def},~\eqref{1-FmathfrakY} and~\cite[eqn.~(3.383-10),~p.~348]{Grad}, the analytical expression for the ergodic rate for symbol $s_2$ in CRS-NOMA with MRC in Nakagami-$m$ fading reduces to~\eqref{C_s2_MRC_Closed}.
\end{proof}
Therefore, using~\eqref{C_s1_MRC_Closed} and~\eqref{C_s2_MRC_Closed}, the ergodic rate for the CRS-NOMA using MRC is given by 
\begin{align}
	\bar C_{\mathrm{MRC}} = \bar C_{s_1, \mathrm{MRC}} + \bar C_{s_2, \mathrm{MRC}}. \label{C_sum_MRC}
\end{align}

It can be shown using some algebraic manipulations that for $m_{sr} = m_{sd} = m_{rd} = 1$, \eqref{C_sum_MRC} reduces to~\cite[eqn.~(16)]{SCC}. Furthermore, for $m_{sr} = m_{sd} = m_{rd} = 1$ and $N_r = N_d = 1$, \eqref{C_sum_MRC} reduces to~\cite[eqn.~(14)]{CRS_NOMA}.

In contrast to this, following a similar line of reasoning as in the previous section for CRS-OMA with SC, the ergodic rate for the case of CRS-OMA with MRC can be given by
\begin{align*}
	\bar C_{\mathrm{MRC-OMA}} = 0.5 \mathbb E_{\mathcal W} \left[ \log_2(1 + \mathcal W \rho)\right],
\end{align*}
where $\mathcal W \triangleq \min\{\mathfrak g_{sr}, \mathfrak g_{sd} + \mathfrak g_{rd}\}$.
%
%
\subsection{High-SNR approximation of the ergodic rate} \label{Sec-MRC_HighSNR}
Similar to the case of CRS-NOMA with SC, a high-SNR ($\rho \gg 1$) approximation of the ergodic rate of $s_1$ in the case of CRS-NOMA with MRC can deduced using~\eqref{C_s1_MRC_Def} via
\begin{align}
	\bar C_{s_1, \mathrm{MRC}} = & \ \dfrac{1}{2} \mathbb E \left\{ \log_2 \left( 1 + \dfrac{a_1 \rho \mathcal X}{a_2 \rho \mathcal X + 1}\right)\right\} \notag \\ 
	 \approx & \ \dfrac{1}{2} \log_2 \left( 1 + \dfrac{a_1}{a_2}\right). \label{C_s1_MRC_HighSNR}
\end{align}
Note that the high-SNR approximation of the ergodic rate of $s_1$ in CRS-NOMA with MRC shown in~\eqref{C_s1_MRC_HighSNR} is independent of $\rho$ and is equal to that of the CRS-NOMA with SC. On the other hand, for the case of $s_2$, it follows from~\eqref{C_s2_MRC_Def} that for $\rho \gg 1$, we have 
\begin{align}
	\bar C_{s_2 \mathrm{MRC}} \approx & \ \dfrac{1}{2} \log_2 \rho + \dfrac{1}{2} \mathbb E \left\{ \log_2 \mathcal Y\right\} \notag \\
	= & \ \dfrac{1}{2} \log_2 \rho + \dfrac{1}{2} \int_0^\infty \log_2(x) f_{\mathcal Y}(x) \mathrm dx. \label{C_s2_MRC_HighSNR_integral}
\end{align}
Using~\eqref{1-FmathfrakY}, the expression for $f_{\mathcal Y}(x)$ can be given by 
\begin{align}
	f_{\mathcal Y}(x) = & \ \dfrac{\mathrm d F_{\mathcal Y}(x)}{\mathrm dx} \notag \\
	= & \sum_{\mu = 0}^{m_{sr} N_r - 1} \sum_{\nu = 0}^{m_{rd}N_d - 1} \dfrac{m_{sr}^\mu m_{rd}^\nu \exp(-\Xi x) }{a_2^{\mu}\Omega_{sr}^\mu \Omega_{rd}^\nu \mu! \nu!} \notag \\
	& \hspace{1.5cm}\times \left[ \Xi x^{\mu + \nu} - (\mu + \nu) x^{\mu + \nu - 1} \right]. \label{f_mathcalY}
\end{align}
Substituting the expression for $f_{\mathcal Y}(x)$ from~\eqref{f_mathcalY} into~\eqref{C_s2_MRC_HighSNR_integral}, and solving the integral using~\cite[eqn.~(4.352-1),~p.~573]{Grad}, we obtain
\begin{align}
	& \bar C_{s_2, \mathrm{MRC}} \notag \\
	\approx  & \dfrac{1}{2} \log_2 \rho +   \dfrac{1}{2 \ln 2}\!\!\sum_{\mu = 0}^{m_{sr} N_r - 1} \sum_{\nu = 0}^{m_{rd}N_d - 1} \!\!\! \dfrac{m_{sr}^\mu m_{rd}^\nu}{a_2^{\mu}\Omega_{sr}^\mu \Omega_{rd}^\nu \mu! \nu! \Xi^{\mu + \nu}} \notag \\
	& \times [ \Gamma(\mu + \nu + 1) \left\{ \psi(\mu + \nu + 1) - \ln \Xi \right\} \notag \\
	& \hspace{1cm}- (\mu + \nu)\Gamma(\mu + \nu) \left\{ \psi(\mu + \nu) - \ln \Xi \right\}], \label{C_s2_MRC_HighSNR}
\end{align}
By adding~\eqref{C_s1_MRC_HighSNR} and~\eqref{C_s2_MRC_HighSNR}, an analytical expression can be obtained for the high-SNR approximation of $\bar C_{\mathrm{MRC}}$.
%
%

Following a similar line of argument as given for the case of CRS-NOMA with SC, it is straightforward to conclude using~\eqref{C_s1_MRC_HighSNR}~--~\eqref{C_s2_MRC_HighSNR} that at high SNR, $\bar C_{\mathrm{MRC}}$ grows as $0.5 \log_2 \rho + \mathscr O(1)$.
%
%
\subsection{Outage probability} \label{Sec-MRC_Outage}
Similar to~Section~\ref{Sec-SC_Outage}, in the case of CRS-NOMA with MRC in Nakagami-$m$ fading, the outage probability is defined as
\begin{align}
	& P_{\mathrm{MRC}} \notag \\
	= & \ 1 - \Pr (\gamma_{sd, \mathrm{MRC}} \geq \theta) \Pr \!\left( \gamma_{sr, \mathrm{MRC}}^{(1)} \geq \theta, \gamma_{sr, \mathrm{MRC}}^{(2)} \geq \theta \right) \notag \\
	= & \ 1 - \{1 - F_{\mathfrak g_{sd}} (\Delta_1)\} \{1 - F_{\mathfrak g_{sr}} (\Delta_2)\} \{1 - F_{\mathfrak g_{rd}} (\Delta_3)\}.  \label{Outage_MRC_Def}
\end{align}
Using~\eqref{Outage_MRC_Def},~\eqref{F_mathfrak_gSR} and~\cite[eqn.~(8.352-1),~p.~899]{Grad}, an analytical expression for the outage probability is given by~\eqref{Outage_MRC_Closed}, shown on the next page.
\begin{figure*}[!t]
\normalsize
\setcounter{MYtempeqncnt}{\value{equation}}
\begin{align}
		P_{\mathrm{MRC}} =  1 - & \left\{ 1 -  \dfrac{\gamma \left( m_{sd} N_d, \frac{m_{sd}}{\Omega_{sd}}\Delta_1\right)}{\Gamma(m_{sd} N_d)}  \right\} \left\{ 1 -  \dfrac{ \gamma \left( m_{sr} N_r, \frac{m_{sr}}{\Omega_{sr}}\Delta_2 \right)}{\Gamma(m_{sr} N_r)} \right\}  \left\{ 1 - \dfrac{ \gamma \left( m_{rd} N_d, \frac{m_{rd}}{\Omega_{rd}}\Delta_3 \right) }{\Gamma(m_{rd} N_d)} \right\}. \label{Outage_MRC_Closed}
\end{align}
\setcounter{equation}{\value{MYtempeqncnt}}
\hrulefill
\end{figure*}
\addtocounter{equation}{1}
%
%
\subsection{Diversity analysis} \label{Sec-MRC_Diversity}
Using~\eqref{T_function},~\eqref{T_function_expansion},~\eqref{Outage_MRC_Def},~\eqref{F_mathfrak_gSR} and~\cite[eqn.~(8.352-1),~p.~899]{Grad}, it can be shown that $F_{\mathfrak g_{sd}}(\Delta_1)$, $F_{\mathfrak g_{sr}}(\Delta_2)$ and $F_{\mathfrak g_{rd}}(\Delta_3)$ decay as $\rho^{-m_{sd} N_d}$, $\rho^{-m_{sr} N_r}$ and $\rho^{-m_{rd} N_d}$, respectively, for $\rho \to \infty$. Then following a similar approach as in~Section~\ref{Sec-SC_Diversity}, it can be shown that the diversity order of CRS-NOMA with MRC is equal to $\min\{m_{sd}N_d, m_{sr} N_r, m_{rd}N_d\}$.
%
%
\section{Results and Discussion} \label{Sec-Results}
In this section, we evaluate the analytically derived results for the ergodic rate and outage probability, and compare them with numerically computed results. Unless stated otherwise, we set $\Omega_{sd} = 1$, $\Omega_{sr} = 10$, $\Omega_{rd} = 2.5$ and $R = 1$ bps/Hz. As discussed in~Section~III-B, the design constraint $a_2 < 1/(1 + \theta)$ implies a valid range $0 < a_2 < 1/(1 + \theta)$, i.e., $0 < a_2 < 0.25$. For every given $\rho$, $m_{sd} = m_{sr} = m_{rd} = m$ and $N_r = N_d = N$ (although the analysis presented in this paper also holds good for the case where $m_{sd} \neq m_{sr} \neq m_{rd}$ and $N_r \neq N_d$), we find the optimal value of $a_2$ that minimizes the outage probability of CRS-NOMA, and then use that optimal value of $a_2$ to find the ergodic rate (unless stated otherwise). The optimization of $a_2$ is performed using a numerical search over the finite set $a_2 \in \{0.01, 0.02, \ldots, 0.24\}$. 
\begin{figure}[t]
\centering
\includegraphics[width = 1\linewidth]{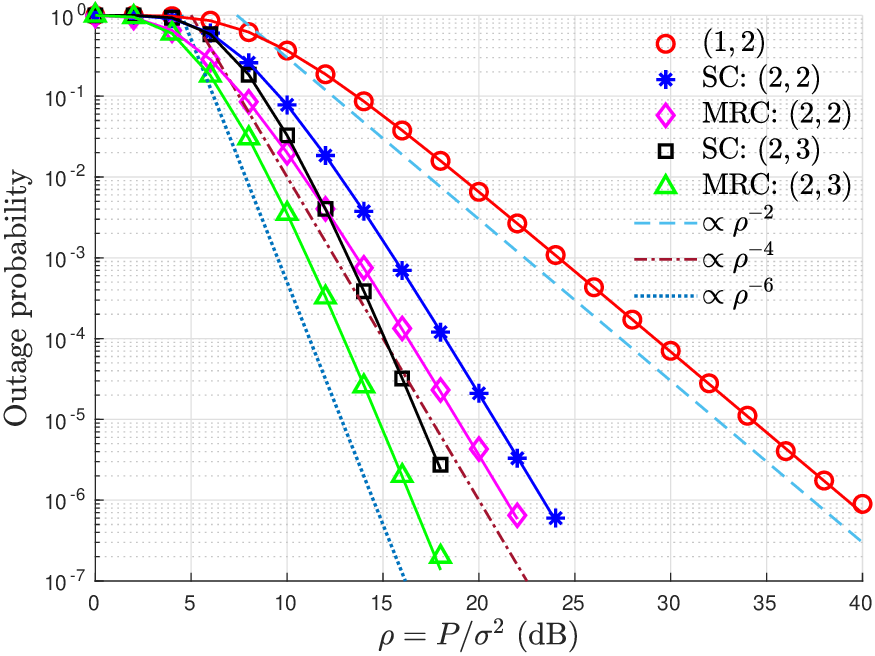}
\caption{Outage probability for CRS-NOMA. Here solid lines represent the analytical results, and the value of $N$ and $m$ are indicated in the form $(N, m)$.}
\label{Outage_Fig}
\end{figure}

\begin{figure}[t]
\centering
  \includegraphics[height = 9cm, width = 8cm]{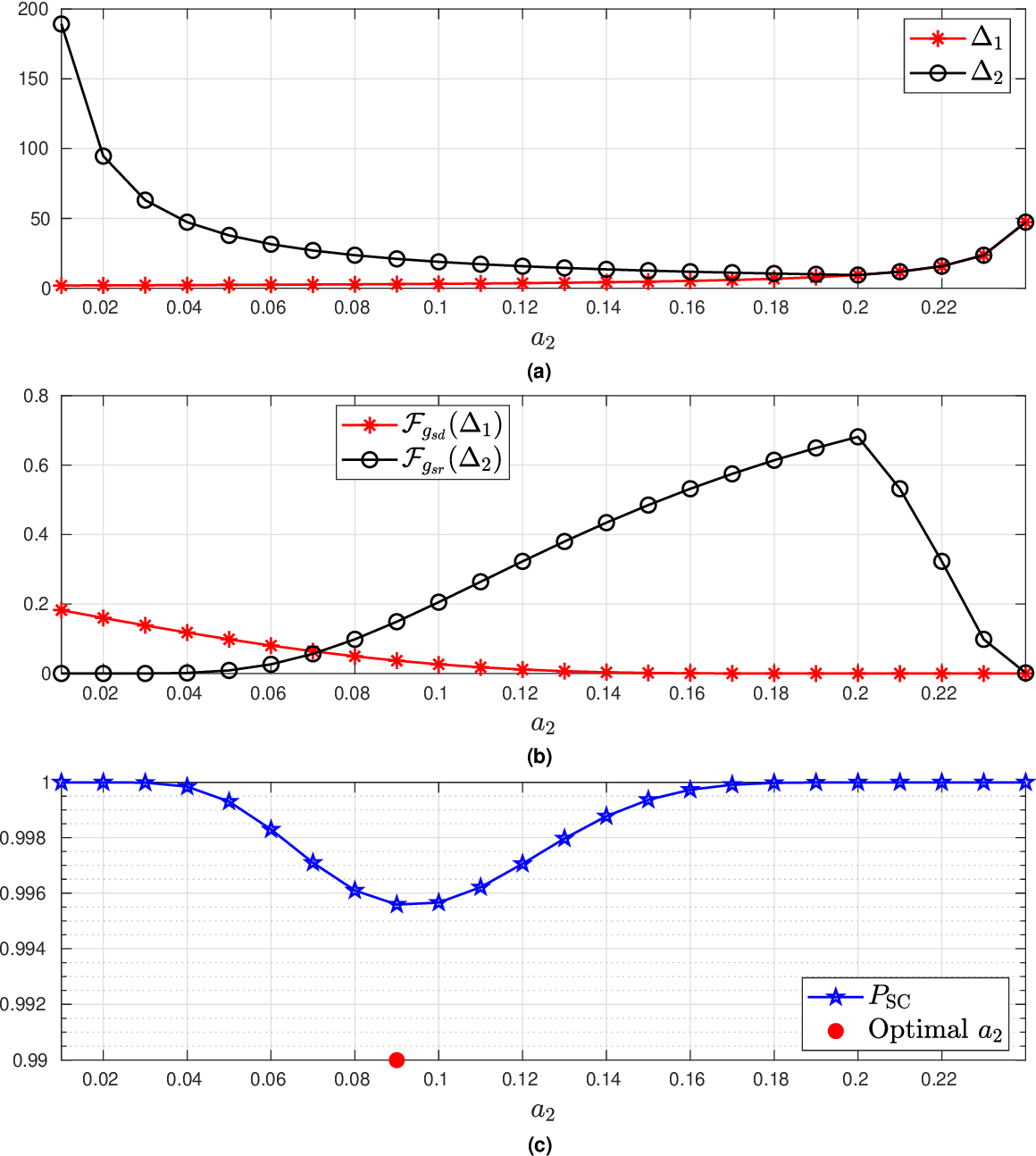}
  \caption{Optimal value of $a_2$ to minimize the outage probability of CRS-NOMA with SC for $m = 2$, $N = 2$ and $\rho = 2$ dB.}
  \label{OutAnalysis2dB}  
\end{figure}

\begin{figure}[t]
  \centering
  \includegraphics[height = 9cm, width = 8cm]{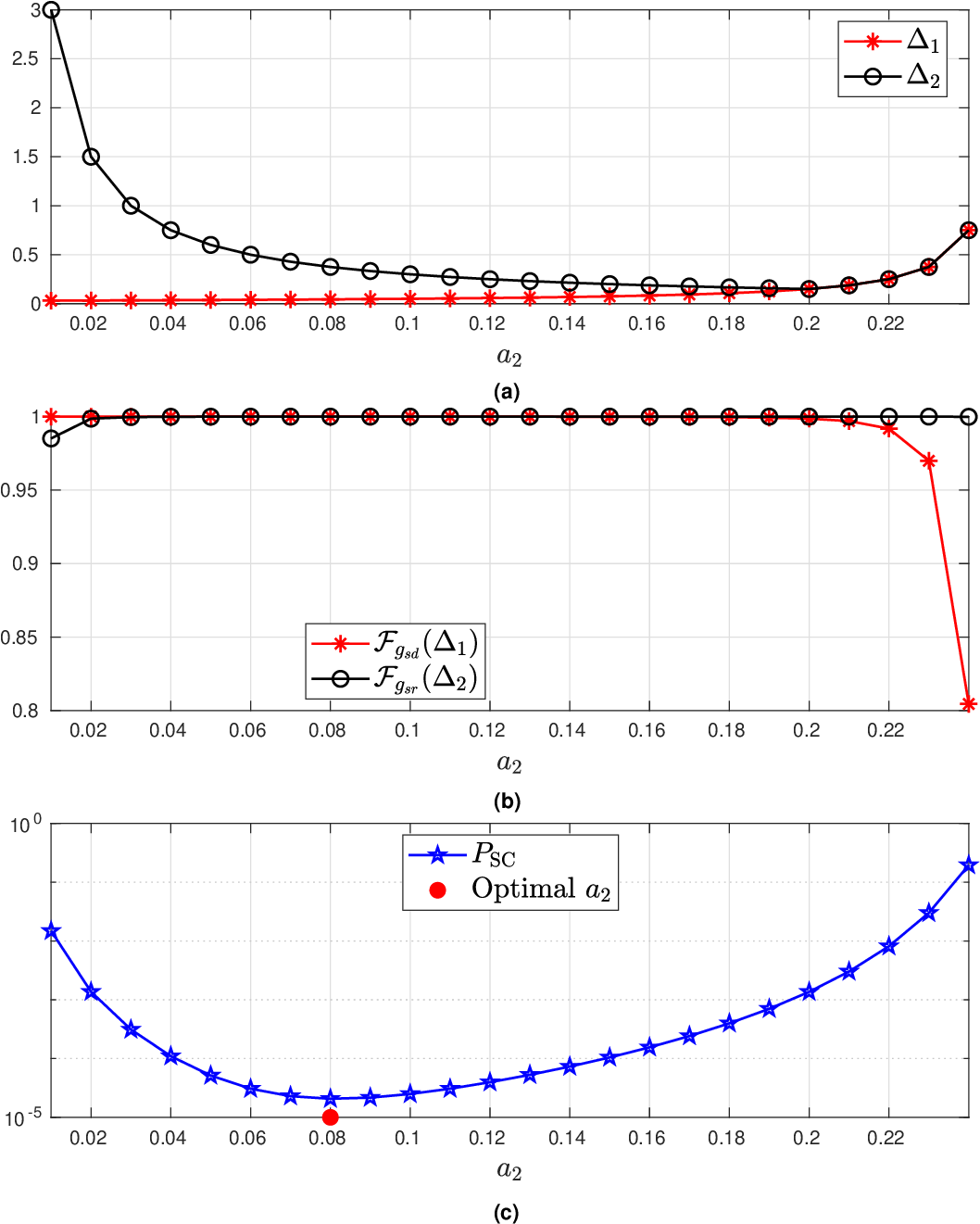}
  \caption{Optimal value of $a_2$ to minimize the outage probability of CRS-NOMA with SC for $m = 2$, $N = 2$ and $\rho = 20$ dB.}
  \label{OutAnalysis20dB}  
\end{figure}
Fig.~\ref{Outage_Fig} shows the outage probability of CRS-NOMA for different values of the shape parameter $m$ and the number of receive antennas $N$. It can be noted from the figure that the outage probability of the system decreases with an increase in the value of the shape parameter as well as the number of receive antennas. In the case of Nakagami-$m$ fading, the amount of fading is inversely proportional to to the shape parameter (the constant of proportionality being equal to 1), therefore, with an increase in the value of $m$, the fading becomes less severe, resulting into a lower outage probability. On the other hand, as the number of antennas $N$ increases, the outage probability decreases by the virtue of the increased spatial diversity. It can also be noted from the figure that the diversity order of CRS-NOMA is equal to $mN$, as derived analytically in~Sections~\ref{Sec-SC_Diversity} and~\ref{Sec-MRC_Diversity}. Therefore, the diversity order depends on the shape parameter $m$ as well as on the number of receive antennas, which is in line with the results reported in~\cite{Branka}.

The optimal value of $a_2$ for the minimization of the outage probability turns out to be between 0.1 and 0.08 for the system parameters selected in this paper. The reason for this is explained via~Figs.~\ref{OutAnalysis2dB} and~\ref{OutAnalysis20dB}. For CRS-NOMA with SC, it can be noted from~\eqref{Outage_SC_Def} that $\Delta_3 (= \theta/\rho)$ is independent of $a_2$. Therefore, the optimal value of $a_2$ to minimize $P_{\mathrm{SC}}$ will depend only on $F_{g_{sd}}(\Delta_1)$ and $F_{g_{sr}}(\Delta_2)$. Denoting $1 - F_{g_{sd}}(\Delta_1)$ and $1 - F_{g_{sr}}(\Delta_2)$ by $\mathcal F_{g_{sd}}(\Delta_1)$ and $\mathcal F_{g_{sr}}(\Delta_2)$, respectively, which are the complementary CDFs (CCDFs) of the random variables $g_{sd}$  and $g_{sr}$, respectively, the product $\mathcal F_{g_{sd}}(\Delta_1) \mathcal F_{g_{sr}}(\Delta_2)$ should be maximized in order to minimize $P_{\mathrm{SC}}$. Fig.~\ref{OutAnalysis2dB}~(a) shows the variation of $\Delta_1$ and $\Delta_2$ w.r.t. $a_2$. It can be noted from the figure that the value of $\Delta_1$ becomes minimal for $a_2 = 0.01$, whereas the value of $a_2$ that minimizes $\Delta_2$ is found to be equal to 0.2. Therefore, $\mathcal F_{g_{sd}}(\Delta_1)$ and $\mathcal F_{g_{sr}}(\Delta_2)$ will attain the corresponding maximum values at $a_2 = 0.01$ and $a_2 = 0.2$, as shown in~Fig.~\ref{OutAnalysis2dB}~(b). Since $P_{\mathrm{SC}}$ is a function of $\mathcal F_{g_{sd}}(\Delta_1)$ and $\mathcal F_{g_{sr}}(\Delta_2)$, and the product $\mathcal F_{g_{sd}}(\Delta_1) \times \mathcal F_{g_{sr}}(\Delta_2)$ attains a maximum value at $a_2 = 0.09$, the optimal value of $a_2$ that minimizes $P_{\mathrm{SC}}$ turns out to be $0.09$, as shown in~Fig.~\ref{OutAnalysis2dB}~(c). Furthermore,~Fig.~\ref{OutAnalysis2dB}~(c) demonstrates that choosing the optimal value of $a_2$ does not have a significant effect on the value of the outage probability, because of the small value of $\rho$ ($= 2$ dB). A similar process to obtain the optimal value of $a_2$ for the minimization of $P_{\mathrm{SC}}$ for $\rho = 20$ dB is shown in~Fig.~\ref{OutAnalysis20dB}, where the effect of the optimal value of $a_2$ can easily be noted (as the values of the outage probability for non-optimal values of $a_2$ are significantly larger). The optimal value of $a_2$ for minimizing $P_{\mathrm{MRC}}$ can be obtained in a similar fashion.

\begin{figure}[t]
\centering
	\includegraphics[width = 1\linewidth]{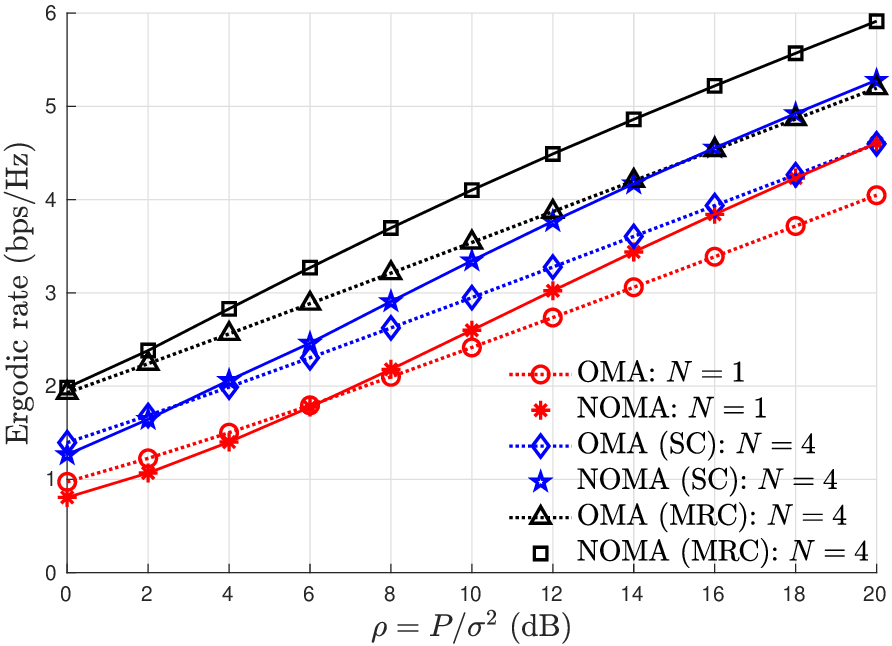}
\caption{Ergodic rate for $m = 2$ and different values of $N$. The analytical results are represented using solid lines.}
\label{RateN}
\end{figure}

\begin{figure}[t]
\centering
  \includegraphics[width = 1\linewidth]{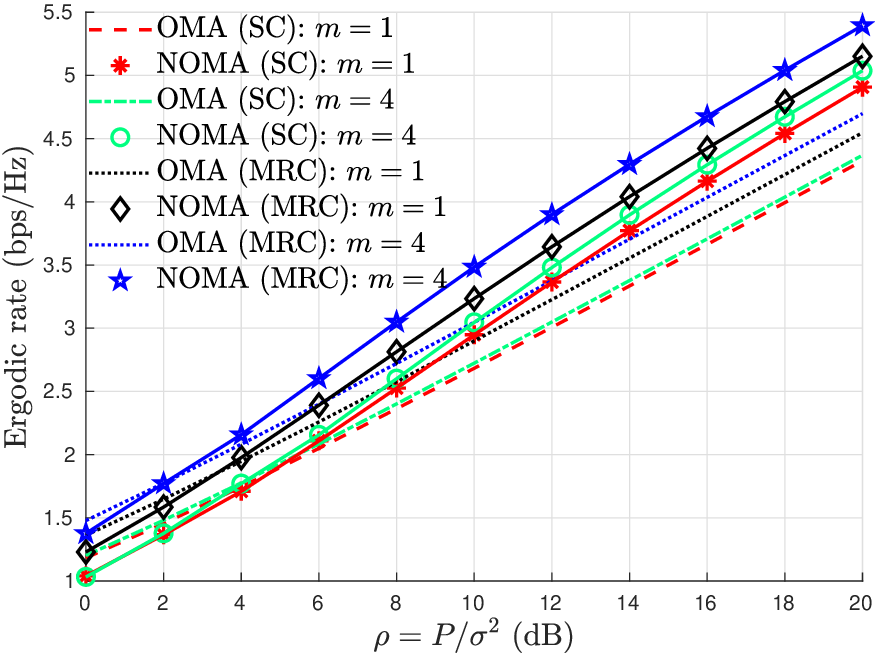}
	\caption{Ergodic rate for $N = 2$ and different values of $m$. The analytical results are represented using solid lines.}
\label{RateM}
\end{figure}

\begin{figure}[t]
\centering
	\includegraphics[width = 1\linewidth]{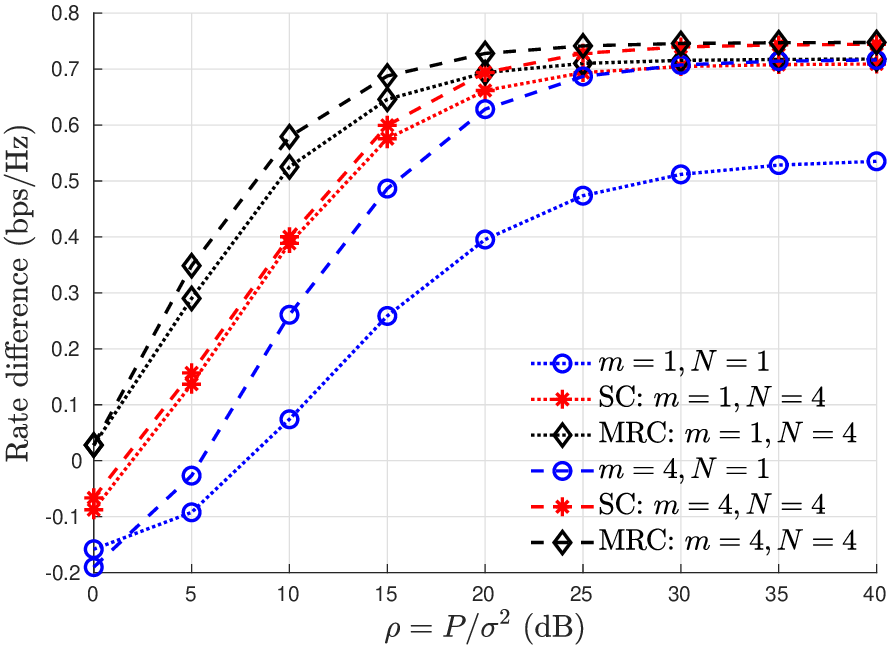}
	\caption{Difference between the ergodic rate for CRS-NOMA and CRS-OMA.}
	\label{RateDiff_fig}
\end{figure}

Fig.~\ref{RateN} shows the effect of $N$ on the ergodic rates for $m = 2$. It can be noticed from the figure that for the single-antenna system ($N = 1$), OMA performs better than NOMA in the low-$\rho$ regime. It is evident from the figure that by using a multi-antenna receiver architecture, CRS-NOMA outperforms CRS-OMA at a lower value of $\rho$. Moreover, it can be noted that MRC reception has a significant advantage over SC reception in terms of the ergodic rate. Fig.~\ref{RateM} shows the effect of the shape parameter $m$ on the ergodic rate for $N = 2$. It can be seen from the figure that as the value of the shape parameter $m$ increases, the value of $\rho$ at which CRS-NOMA outperforms CRS-OMA decreases, because of less severe fading. The advantage of using MRC receivers over SC receivers are also clearly visible from the figure.

Fig.~\ref{RateDiff_fig} shows the effect of the shape parameter $m$ and the number of antennas $N$ on the difference between the ergodic rate of the CRS-NOMA and its OMA-based counterpart. It is evident from the figure that the performance superiority of CRS-NOMA over CRS-OMA increases with an increase in the number of antennas and/or the shape parameter $m$. It is also evident from the figure that this performance difference increases monotonically in the low-to-mid $\rho$ regime and then saturates for large $\rho$. Note that we have proved that the ergodic rate of the CRS-NOMA grows as $0.5 \log_2 \rho$ for large values of $\rho$ (irrespective of the value of $m$ and $N$), and using the arguments in~\cite[p.~37]{MIMO_Goldsmith}, it can be proved that the ergodic rate of the CRS-OMA also grows as $0.5 \log_2 \rho$ for large $\rho$. Since the ergodic rate for both CRS-NOMA and CRS-OMA have the same slope for large $\rho$, this is the reason that the difference between the ergodic rate of CRS-NOMA and CRS-OMA saturates at high SNR.

\begin{figure}[t]
  \centering
  \includegraphics[width = 1\linewidth]{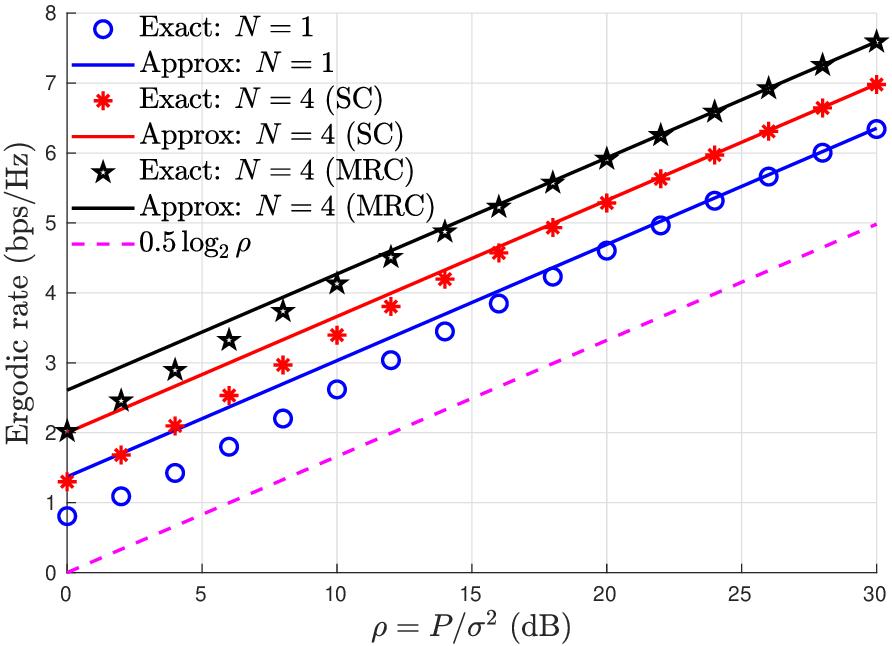}
	\caption{Comparison of the exact ergodic rate and the corresponding high-SNR approximation, for $m = 2$ and fixed $a_2 = 0.1$.}
	\label{HighSNRfig}
\end{figure}

Finally, Fig.~\ref{HighSNRfig} shows a comparison of the exact ergodic rate (plotted via numerical evaluation) and the corresponding high-SNR approximation (plotted using the analytical expression). A tight agreement between the exact and the approximate results in the mid-to-high SNR regime confirms the accuracy of the analysis. It is also noteworthy that the ergodic rate becomes parallel to $0.5 \log_2 \rho$ in the high-SNR regime, which verifies the results established in Sections~\ref{Sec-SC_HighSNR} and~\ref{Sec-MRC_HighSNR}, and also indicates that the slope of the ergodic rate at high SNR does not depend on the number of receive antennas or the value of the Nakagami shape parameter.
%
%
\section{Conclusion} \label{Sec-Conclusion}
In this paper, we presented a comprehensive performance analysis of the multiple-antenna-assisted non-orthogonal multiple access-based cooperative relaying system (CRS-NOMA) in the presence of Nakagami-$m$ fading, considering two different signal combining schemes -- SC and MRC. In particular, we have derived exact analytical expressions for the ergodic rate and outage probability, along with a high-SNR approximation of the ergodic rate. Regarding the ergodic rate, our results show that in contrast to existing CRS-NOMA systems, the CRS-NOMA with receive diversity outperforms the corresponding OMA-based system even in the region of low transmit SNR. We also provide an explicit analytical proof that the CRS-NOMA with receive diversity achieves full diversity order, which depends on the number of antennas as well as the Nakagami-$m$ shape parameter. The results also indicate that increasing the number of receive antennas and/or the value of $m$ also increases the performance gap between CRS-NOMA and CRS-OMA (especially in the region of low transmit SNR). For any fixed values of number of receive antennas and shape parameter, this performance difference increases monotonically with an increase in the transmit SNR regime and saturates in the region of large transmit SNR. We also provided an explicit analytical proof that at high SNR, the ergodic rate grows as $0.5 \log_2 \rho$ for both SC and MRC receivers. Due to the generality of the Nakagami-$m$ fading model and the multiplicity of diversity combining schemes considered, the results obtained in this paper serve as a practical system design tool and can be used to analyze the performance of the CRS-NOMA with receive diversity under a variety of fading models of practical interest. Some interesting directions to further enhance the spectral efficiency of the CRS-NOMA include the consideration of full-duplex relaying and multiple-input multiple-output (MIMO) communication. Also, investigation of the performance of CRS-NOMA considering imperfect SIC, correlated fading links, imperfect/outdated CSI and finite blocklength coding represent further interesting directions for future research.
\appendices
\section{Proof of Theorem~\ref{Theorem-SC_Cs1}} \label{Proof-SC_Cs1}
Using a transformation of random variables and assuming $m_{sr} \in \mathbb Z_+$, the CDF of $g_{sr}$ can be given by~(c.f. \cite[eqn.~(5)]{Yacoub},~\cite[eqn.~(8.352-1),~p.~899]{Grad})
\begin{align}
	& F_{g_{sr}}(x) = \left[ \dfrac{\gamma \left( m_{sr}, \frac{m_{sr}}{\Omega_{sr}} x\right)}{\Gamma(m_{sr})} \right]^{N_r} \notag \\
	= & \ \left[ 1 - \exp \left( \dfrac{-m_{sr}x}{\Omega_{sr}} \right) \sum_{\mu = 0}^{m_{sr} - 1} \dfrac{m_{sr}^{\mu} x^{\mu}}{\Omega_{sr}^{\mu} \mu!} \right]^{N_r} \notag \\
	= & \ 1 + \sum_{\substack{k_0 + k_1 + \cdots + k_{m_{sr}} = N_r \\ k_0 \neq N_r}} \binom{N_r}{k_0, k_1, \ldots, k_{m_{sr}}} (-1)^{N_r - k_0} \notag \\
	& \times \!\left[ \prod_{\mu = 0}^{m_{sr} - 1} \left(\! \dfrac{1}{\mu!} \!\right)^{\!k_{\mu + 1}}\!\right]\!\left(\! \dfrac{m_{sr}}{\Omega_{sr}}\! \right)^{\!\!\tau} \!x^\tau \exp \left[\dfrac{-(N_r - k_0)m_{sr}}{\Omega_{sr}}x \right], \label{F_gSR_SC}
\end{align}
where $\tau = \sum_{\mu = 0}^{m_{sr} - 1} \mu k_{\mu + 1}$. Similarly, for $m_{sd} \in \mathbb Z_+$, we have
\begin{align}
	& F_{g_{sd}}(x) = 1 + \sum_{\substack{{l_0 + l_1 + \cdots + l_{m_{sd}} = N_d} \\ l_0 \neq N_d}} \!\!\!\binom{N_d}{l_0, l_1, \ldots, l_{m_{sd}}} (-1)^{N_d - l_0} \notag \\
	& \times \left[ \prod_{\nu = 0}^{m_{sd} - 1} \left( \dfrac{1}{\nu!}\right)^{l_{\nu + 1}}\right]\left( \dfrac{m_{sd}}{\Omega_{sd}}\right)^{\omega}x^\omega \exp \left[\dfrac{-(N_d - l_0)m_{sd}}{\Omega_{sd}}x \right], \notag
\end{align}
where $\omega = \sum_{\nu = 0}^{m_{sd} - 1} \nu l_{\nu + 1}$ and $\gamma(\cdot, \cdot)$ denotes the lower-incomplete Gamma function.
\begin{figure*}[!t]
\normalsize
\setcounter{MYtempeqncnt}{\value{equation}}
\begin{align}
	& 1 - F_{X} (x) = 1 - F_{g_{sr}}(x) - F_{g_{sd}}(x) + F_{g_{sr}}(x)F_{g_{sd}}(x) \notag \\
	 = &  \!\!\!\! \sum_{\substack{{k_0 + k_1 + \cdots + k_{m_{sr}} = N_r} \\ k_0 \neq N_r}}  \ \sum_{\substack{{l_0 + l_1 + \cdots + l_{m_{sd}} = N_d} \\ l_0 \neq N_d}} \binom{N_r}{k_0, k_1, \ldots, k_{m_{sr}}} \binom{N_d}{l_0, l_1, \ldots, l_{m_{sd}}} (-1)^{N_r + N_d - k_0 - l_0} \notag \\
		\times &  \left[ \prod_{\mu = 0}^{m_{sr} - 1} \left( \dfrac{1}{\mu!}\right)^{k_{\mu + 1}}\right] \left( \dfrac{m_{sr}}{\Omega_{sr}}\right)^{\tau} \left[ \prod_{\nu = 0}^{m_{sd} - 1} \left( \dfrac{1}{\nu!}\right)^{l_{\nu + 1}}\right]\left( \dfrac{m_{sd}}{\Omega_{sd}}\right)^{\omega}x^{\tau + \omega} \exp \left[ - \underbrace{\left\{ \dfrac{(N_r - k_0) m_{sr}}{\Omega_{sr}} + \dfrac{(N_d - l_0) m_{sd}}{\Omega_{sd}} \right\}}_{\Psi_{k_0, l_0}} x\right]. \label{1-FmathscrX}
\end{align}
\setcounter{equation}{\value{MYtempeqncnt}}
\hrulefill
\end{figure*}
\addtocounter{equation}{1}
Therefore, the expression for $1 - F_{ X}(x)$ can be given by~\eqref{1-FmathscrX}, shown at the top of the page. Using~\eqref{C_s1_SC_Integral_Nak},~\eqref{1-FmathscrX} and~\cite[eqn.~(3.383-10),~p.~348]{Grad}, the analytical expression for $\bar C_{\mathrm{s_1, \mathrm{SC}}}$ reduces to \eqref{C_s1_SC_Closed_Nak}; this completes the proof.
%
%
\section{Proof of Theorem~\ref{Theorem-MRC_Cs1}} \label{Proof-MRC_Cs1}
Using a transformation of random variables, the probability density function (PDF) of $\mathfrak g_{sr}(x)$ is given by~(c.f.~\cite[eqn.~(2)]{Branka})
\begin{align}
	f_{\mathfrak g_{sr}}(x) = \dfrac{x^{m_{sr}N_r - 1} m_{sr}^{m_{sr}N_r}}{\Omega_{sr}^{m_{sr}N_r} \Gamma(m_{sr, N_r})} \exp \left( \dfrac{-m_{sr}x}{\Omega_{sr}} \right). \notag 
\end{align}
Therefore, using~\cite[eqn.~(3.381-1),~p.~346]{Grad} and~\cite[eqn.~(8.351-1),~p.~899]{Grad}, the CDF of $\mathfrak g_{sr}$ (for $m_{sr} \in \mathbb Z_+$) is given by 
\begin{align}
	F_{\mathfrak g_{sr}}(x) = & \ \int_{0}^{x}f_{ \mathfrak g_{sr}}(t) \ \mathrm dt \notag \\
	  = & \ 1 - \exp \left( \dfrac{-m_{sr}x}{\Omega_{sr}}\right) \sum_{\mu = 0}^{m_{sr}N_r - 1} \dfrac{m_{sr}^\mu x^\mu}{\Omega_{sr}^\mu \mu!}. \label{F_mathfrak_gSR}
\end{align}
Similarly, the CDF of $\mathfrak g_{sd}(x)$ is given by
\begin{align}
	F_{\mathfrak g_{sd}}(x) = 1 - \exp \left( \dfrac{-m_{sd}x}{\Omega_{sd}} \right) \sum_{\nu = 0}^{m_{sd}N_d - 1} \dfrac{m_{sd}^\nu x^\nu}{\Omega_{sd}^\nu \nu!}. \notag
\end{align}
Therefore, for $\mathcal X = \min\{\mathfrak g_{sr}, \mathfrak g_{sd}\}$, 
\begin{align}
	& 1 - F_{\mathcal X}(x) = 1 - F_{\mathfrak g_{sr}}(x) - F_{\mathfrak g_{sd}}(x) + F_{\mathfrak g_{sr}}(x)  F_{\mathfrak g_{sd}}(x) \notag \\
	= &\ \exp(-\Phi x) \sum_{\mu = 0}^{m_{sr} N_r - 1} \sum_{\nu = 0}^{m_{sd}N_d - 1} \dfrac{m_{sr}^\mu m_{sd}^\nu x^{\mu + \nu}}{\Omega_{sr}^\mu \Omega_{sd}^\nu \mu! \nu!}, \label{1-FmathfrakX}
\end{align}
where $\Phi = (m_{sr}/\Omega_{sr}) + (m_{sd}/\Omega_{sd})$. Using~\eqref{C_s1_MRC_Def},~\eqref{1-FmathfrakX} and~\cite[eqn.~(3.383-10),~p.~348]{Grad}, the analytical expression for $\bar C_{s_1, \mathrm{MRC}}$ reduces to~\eqref{C_s1_MRC_Closed}; this completes the proof.
\balance
\bibliographystyle{IEEEtran}
\bibliography{open}

\end{document}